\documentclass[pdflatex,sn-mathphys]{sn-jnl}

\usepackage{pdfpages}

\jyear{2021}%

\theoremstyle{thmstyleone}%
\theoremstyle{thmstyletwo}%

\theoremstyle{thmstylethree}%

\begin{document}

\title[Article Title]{Bag of States: A Non-sequential Approach to Video-based Engagement Measurement}


\author*[1]{\fnm{Ali} \sur{Abedi}}\email{ali.abedi@uhn.ca}

\author[2]{\fnm{Chinchu} \sur{Thomas}}\email{chinchu.thomas@iiitb.ac.in}

\author[2]{\fnm{Dinesh} \sur{Babu Jayagopi}}\email{jdinesh@iiitb.ac.in}

\author[1]{\fnm{Shehroz} \sur{S. Khan}}\email{shehroz.khan@uhn.ca}

\affil[1]{\orgdiv{KITE}, \orgname{University Health Network, Canada}}

\affil[2]{\orgdiv{Multimodal Perception Lab}, \orgname{ International Institute of Information Technology Bangalor, India}}


\abstract{Automatic measurement of student engagement provides helpful information for instructors to meet learning program objectives and individualize program delivery. Students' behavioral and emotional states need to be analyzed at fine-grained time scales in order to measure their level of engagement. Many existing approaches have developed sequential and spatiotemporal models, such as recurrent neural networks, temporal convolutional networks, and three-dimensional convolutional neural networks, for measuring student engagement from videos. These models are trained to incorporate the order of behavioral and emotional states of students into video analysis and output their level of engagement. In this paper, backed by educational psychology, we question the necessity of modeling the order of behavioral and emotional states of students in measuring their engagement. We develop bag-of-words-based models in which only the occurrence of behavioral and emotional states of students is modeled and analyzed and not the order in which they occur. Behavioral and affective features are extracted from videos and analyzed by the proposed models to determine the level of engagement in an ordinal-output classification setting. Compared to the existing sequential and spatiotemporal approaches for engagement measurement, the proposed non-sequential approach improves the state-of-the-art results. According to experimental results, our method significantly improved engagement level classification accuracy on the IIITB Online SE dataset by 26\% compared to sequential models and achieved engagement level classification accuracy as high as 66.58 \% on the DAiSEE student engagement dataset.}


\maketitle

\section{Introduction}
\label{sec:introduction}
The widespread availability and use of internet services have resulted in a significant increase in the prevalence and mainstream acceptance of virtual learning programs \cite{mukhtar2020advantages}. Compared to traditional in-person learning programs, virtual learning provides many advantages in terms of accessibility, affordability, and customizability. However, virtual learning programs also present other types of challenges \cite{dung2020advantages}, such as putting students and tutors behind a "virtual wall", which could hinder student-tutor interaction and make it difficult for the tutor to assess student engagement. The problem is further aggravated in large groups of students \cite{sumer2021multimodal}. From the tutor's perspective, it is essential to assess student engagement to provide real-time feedback and take appropriate actions to help students achieve their learning goals \cite{gray2016effects}.

Sinatra et al. \cite{sinatra2015challenges} stated that in engagement measurement, the focus is on the student's behavioral, affective, and cognitive states at the moment of interaction with a specific context. Engagement is not stable over time and is best assessed using physiological and psychological measurements at fine-grained time scales \cite{sinatra2015challenges,woolf2009affect,d2012dynamics}. Behavioral engagement entails general on-task behavior and surface-level attention. The indicators of behavioral engagement in the moment of interaction include eye contact, blink rate, and head posture \cite{d2017advanced,woolf2009affect}. Affective engagement is defined as a student's emotional or affective response to the context. Positive versus negative and activating versus deactivating emotions are indicators of affective engagement \cite{sinatra2015challenges,d2017advanced,woolf2009affect}. Cognitive engagement involves the student's psychological commitment and effort allocation to thoroughly comprehend the learning materials \cite{sinatra2015challenges}. Cognitive engagement can be assessed by analyzing information such as student speech in order to recognize the degree to which the student comprehends the context \cite{sinatra2015challenges}. As opposed to behavioral and affective engagements, measuring cognitive engagement requires awareness of the context \cite{bosch2016detecting,sinatra2015challenges}. Measuring students' engagement in a specific context depends on the knowledge about the student and the context. From the standpoint of data analysis, it depends on the data modalities available to analyze. Our paper focuses on the automatic measurement of engagement based on the video. The only data modality is video, without audio, and with no knowledge about the context.

\begin{figure}
    \centering
    \includegraphics[scale=.3]{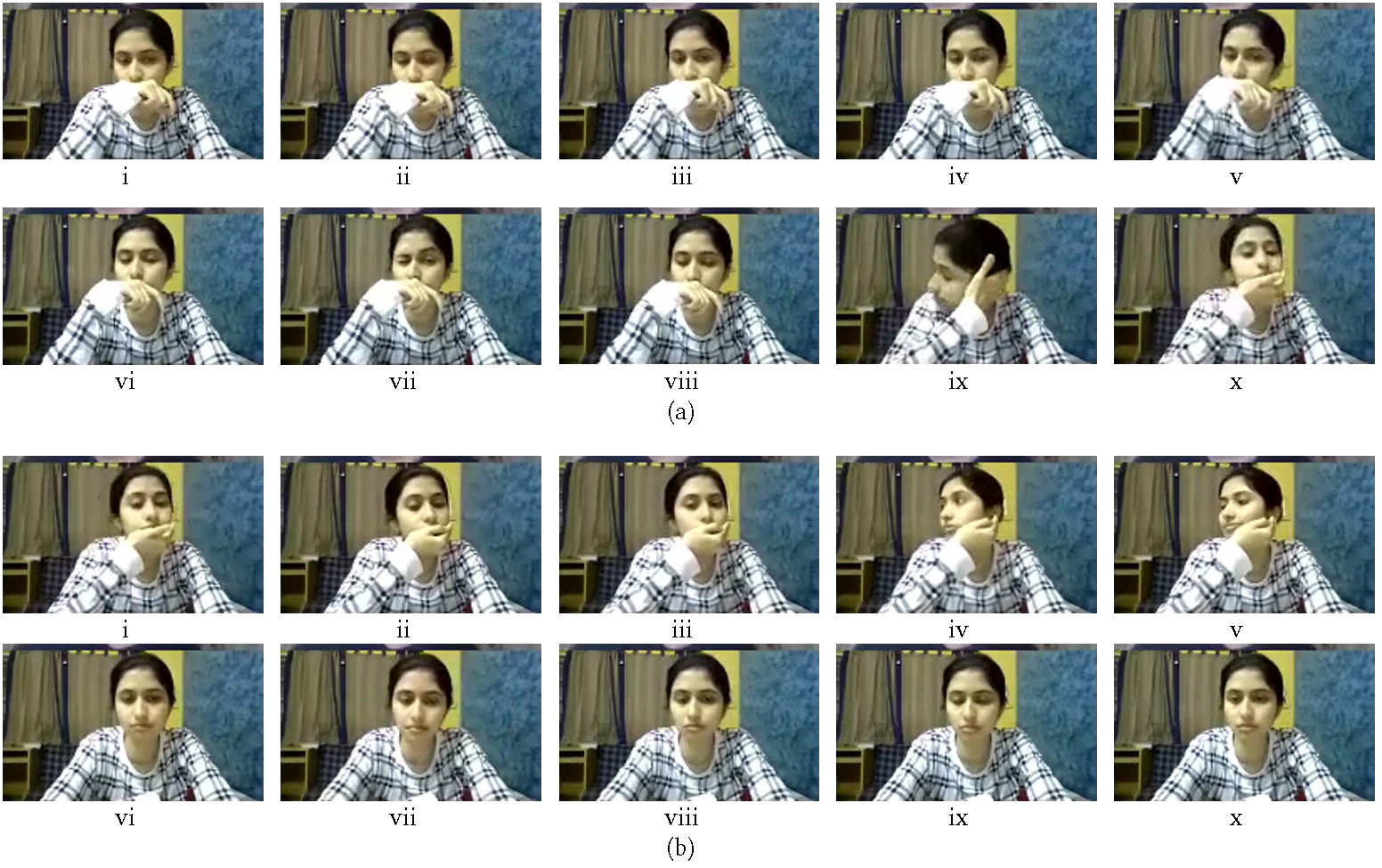}\\
    \caption{
    10 (out of 2400) equally distanced frames from two exemplary video samples of a student in the IIITB Online SE dataset \cite{thomas2022automatic}, (a) and (b). In both (a) and (b), the frames i, ii, ..., x have frame indices of 240, 480, ..., 2400, respectively. There is facepalming in all the video frames in (a) and in half of the video frames in (b). While an off-task head pose distraction occurs in the second half of the video in (a), it occurs in the first half of the video in (b). Eye closure and sleepiness (very low arousal) occur in both the first and second halves in (a) but only in the second half in (b). Despite these differences in the order of the behavioral and affect states of the student in the videos in (a) and (b), the student was annotated as not-engaged in both of the videos. Being not-engaged can be measured based on only the occurrence of the above-mentioned behavioral and affect states and not the order in which they occur.
    }
    \label{fig:frames}
\end{figure}

Most recent works on measuring student engagement have used video data acquired by cameras/webcams and analyzed using computer-vision, machine-learning, and deep-learning techniques \cite{karimah2022automatic,dewan2019engagement}. The video-based approaches for engagement measurement can be categorized into end-to-end and feature-based approaches. In the end-to-end approaches, raw frames of video are fed to deep-learning 3D Convolutional Neural Network (CNN) models (which are capable of analyzing videos) or a combination of 2D CNNs (which can analyze frames of videos) and temporal/sequential models such as Recurrent Neural Networks (RNNs) and Temporal Convolutional Networks (TCNs) \cite{abedi2021improving,liao2021deep,hu2022optimized,gupta2016daisee,zhang2019novel}. In the feature-based approaches, first, handcrafted behavioral and affective features are extracted from consecutive video frames or video segments and then analyzed by temporal/sequential models such as RNNs and TCNs \cite{copur2022engagement,whitehill2014faces,liao2021deep,niu2018automatic,thomas2018predicting,huang2019fine,fedotov2018multimodal,chen2019faceengage,booth2017toward,kaur2018prediction,wu2020advanced,ma2021automatic}. Therefore, currently available end-to-end and feature-based approaches attempt to model the temporal changes in behavioral and affective states of students, as well as the order in which the changes occur, and output the level of engagement. However, none of the existing definitions of student engagement or the protocols for engagement annotation have specified that the behavioral and affective states of a student must occur in a particular order before the student is considered to be at a particular level of engagement \cite{ocumpaugh2015baker,aslan2017human,sinatra2015challenges,khan2022inconsistencies}, see Figure \ref{fig:frames} for an example and discussion about the order of behavioral and affective states.

This paper questions the necessity of modeling the order of behavioral and affective states of students and temporal changes in the states of students in measuring their level of engagement. Accordingly, for the first time in the field of engagement measurement, instead of using the temporal/sequential models, we develop Bag-of-Words (BoW)-based non-sequential models \cite{galke2022bag,wang2013bag}. We call our approach Bag-of-States (BoS) which analyzes only the occurrence of behavioral and affective states and not the order of their occurrences. Various behavioral and affect features are extracted from videos and analyzed by ordinal-output BoS models to output the level of engagement. Our experimental results on two video-based engagement measurement datasets indicated that the lightweight BoS models, requiring less data to be generalized, have superior performance in engagement measurement compared to the sequential/temporal models.

The paper is structured as follows. Section \ref{sec:related_work} describes related works on video-based engagement measurement. Section \ref{sec:is_order_important} discusses the necessity of incorporating the order of affect and behavioral states in engagement measurement. In Section \ref{sec:method}, the proposed pathway for video-based engagement measurement is presented. Section \ref{sec:experiments} describes the experimental settings and results on the proposed methodology. In the end, Section \ref{sec:conclusion} presents our conclusions and directions for future works.

\begin{table}
\centering
\caption{
End-to-end video-based engagement measurement approaches and their deep-learning models, with an explanation of how they analyze sequential video data for engagement measurement, NA: Not Analyzed.
}
\resizebox{\columnwidth}{!}{
    \begin{tabular}{p{.125\linewidth}p{.45\linewidth}p{.25\linewidth}}
 \hline
 Ref., Year & Model & Sequential Data Analysis Approach \\
 \hline\hline
 \cite{gupta2016daisee}, 2016 & C3D & 3D Convolutions\\ \hline
 \cite{zhang2019novel}, 2019 & Inflated 3D CNN (I3D) & 3D Convolutions\\ \hline
 \cite{abedi2021improving}, 2021 & C3D & 3D Convolutions\\ \hline
 \cite{mehta2022three}, 2022 & 3D DenseNet & 3D Convolutions\\ \hline
 \cite{gupta2016daisee}, 2016 & 2D CNN + LSTM & RNN\\ \hline
 \cite{abedi2021improving}, 2021 & 2D ResNet + LSTM/TCN & RNN/TCN\\ \hline
 \cite{liao2021deep}, 2021 & 2D SE-ResNet + LSTM with attention & RNN with attention\\ \hline
 \cite{selim2022students}, 2022 & 2D EfficientNet + LSTM/TCN & RNN/TCN\\ \hline
 \cite{gupta2016daisee}, 2016 & 2D InceptionNet & NA (Image-based)\\ \hline
 \cite{mohamad2019automatic}, 2019 & 2-layer 2D CNN & NA (Image-based)\\ \hline
 \cite{hu2022optimized}, 2022 & 2D ShuffleNet v2 & NA (Image-based)\\ \hline
\end{tabular}
}
\label{tab:literature_review_end_to_end}
\end{table}

\begin{table}
\centering
\caption{
Feature-based video engagement measurement approaches, their features, and their machine-learning/deep-learning models, with an explanation of how they analyze sequences of features for engagement measurement, NA: Not Analyzed.
}
\resizebox{\columnwidth}{!}{
    \begin{tabular}{p{.125\linewidth}p{.5\linewidth}p{.3\linewidth}p{.25\linewidth}}
 \hline
 Ref., Year & Features & Model & Sequential Data Analysis Approach \\
 \hline\hline
 \cite{whitehill2014faces}, 2014 & box filter, Gabor, AU & Gentle-Boost, SVM, and LR & NA (Functional-based)\\ \hline
 \cite{booth2017toward}, 2017 & Facial Landmarks, AU, Optical Flow, Head Pose & SVM, KNN, RF & NA (Functional-based)\\ \hline
 \cite{kaur2018prediction}, 2018 & LBP-TOP & fully-connected neural network & NA (Functional-based)\\ \hline
 \cite{fedotov2018multimodal}, 2018 & Body Pose, Facial Embedding, Eye Gaze, Speech Features & LR & NA (Functional-based)\\ \hline
 \cite{chen2019faceengage}, 2019 & Gaze Direction, Blink Rate, Head Pose, Facial Embedding & AdaBoost, SVM, KNN, RF & NA (Functional-based)\\ \hline
 \cite{chen2019faceengage}, 2019 & Gaze Direction, Blink Rate, Head Pose, Facial Embedding & IRNN, GRU, LSTM & RNN\\ \hline
 \cite{wu2020advanced}, 2020 & Gaze Direction, Head pose, Body Pose, C3D& LSTM, GRU & RNN\\ \hline
 \cite{ma2021automatic}, 2021 & Gaze Direction, Head Pose, AU, C3D & Neural Turing Machine & RNN\\ \hline
 \cite{thomas2018predicting}, 2018 & Gaze Direction, Head Pose, AU & Dilated TCN & TCN\\ \hline
 \cite{copur2022engagement}, 2022 & Eye Gaze, Head Pose, Head Rotation, AU & LSTM & RNN\\ \hline
 \cite{abedi2021affect}, 2022 & Valence, Arousal, Blink Rate, Gaze Direction, Head Pose, Hand Pose & Ordinal LSTM/TCN & RNN/TCN\\ \hline
 \cite{abedi2022detecting}, 2022 & Valence, Arousal, Blink Rate, Gaze Direction, Head Pose, Hand Pose & TCN Auto-encoder & TCN\\ \hline
 \cite{thomas2022automatic}, 2022 & AU, Micro-Macro-Motion & TCN & TCN\\ \hline
 \textbf{Proposed} & \textbf{Valence, Arousal, Blink Rate, Gaze Direction, Head Pose, Hand Pose} & \textbf{BoW} & \textbf{NA}\\ \hline
\end{tabular}
}
\label{tab:literature_review_end_to_end_feature_based}
\end{table}

\section{Related Work}
\label{sec:related_work}
A summary of existing video-based end-to-end, and feature-based engagement measurement approaches are presented in Tables \ref{tab:literature_review_end_to_end}, and \ref{tab:literature_review_end_to_end_feature_based}, respectively. In the end-to-end approaches, consecutive raw video frames are fed to spatio-temporal 3D CNNs or a combination of 2D CNNs and sequential networks, including RNNs (such as vanilla RNN, Long Short-Term Memory (LSTM), and Gated Recurrent Unit (GRU)), TCNs, and transformers. In the former methods, 3D CNNs utilize 3D convolutions to analyze the spatial information in the frames and the temporal changes between consecutive frames, thereby determining the engagement level. The 3D CNNs used for the purpose of engagement measurement were C3D \cite{gupta2016daisee,abedi2021improving}, Inflated 3D CNN (I3D) \cite{zhang2019novel}, and 3D DenseNet \cite{mehta2022three}. In the latter methods, 2D CNNs and sequential models are jointly trained to automatically extract visual features from raw video frames, analyze the sequences of extracted features, and output the engagement level. Various methods have been explored to this end, Gupta et al. \cite{gupta2016daisee} used Long-Term Recurrent Convolutional Network (LRCN), a combination of 2D CNN and LSTM, Abedi and Khan \cite{abedi2021improving} used 2D ResNet along with TCN, Liao et al. \cite{liao2021deep} used Squeeze-and-Excitation ResNet along with LSTM with global attention, and Selim et al. \cite{selim2022students} used EfficientNet along with LSTM and TCN.

In some other end-to-end approaches, single frames of videos are analyzed by 2D CNNs to measure the level of engagement in the video. Some common models used in this approach include InceptionNet \cite{gupta2016daisee}, ShuffleNet v2 \cite{hu2022optimized}, and 2-layer 2D CNN \cite{mohamad2019automatic}. However, in these end-to-end methods, the importance of measuring engagement at fine-grained time scales \cite{sinatra2015challenges} is overlooked. Furthermore, finding the most representative frame \cite{ren2020best} representing the student's engagement throughout the entire video snippet is challenging in such methods.

In feature-based approaches, feature extraction from video frames is separated from sequential modeling. After handcrafted behavioral and affective features or conventional computer-vision features are extracted from consecutive video frames, sequential models such as RNNs and TCNs are then trained to analyze the sequences of extracted features. \cite{copur2022engagement,whitehill2014faces,liao2021deep,niu2018automatic,thomas2018predicting,huang2019fine,fedotov2018multimodal,chen2019faceengage,booth2017toward,kaur2018prediction,wu2020advanced,ma2021automatic}. Table \ref{tab:literature_review_end_to_end_feature_based} shows that some of the previous methods use conventional computer-vision features such as box filters, Gabor \cite{whitehill2014faces}, LBP-TOP \cite{kaur2018prediction}, and C3D \cite{wu2020advanced}. In some other methods, \cite{chen2019faceengage,abedi2021affect}, a CNN with convolutional layers followed by fully-connected layers is trained on a facial expression recognition dataset to extract facial embedding features. In most of the previous methods, facial Action Units (AUs), eye movement, gaze direction, and head pose features are extracted using OpenFace \cite{baltrusaitis2018openface}, or body pose features are extracted using OpenPose \cite{cao2017realtime}. Sequential models then analyze different concatenations of the above-extracted features to output the engagement level in the video.

In some other feature-based approaches, the extracted features from consecutive video frames are not analyzed as a sequence. Instead, functionals of the extracted sequences of features from each video snippet, such as mean and standard deviation, are calculated to create one single feature vector for each video snippet. The feature vector is then fed to a non-sequential model, such as a Support Vector Machine (SVM), K-Nearest Neighbors (KNN), Logistic Regression (LR), or Random Forest (RF) for engagement level measurement \cite{whitehill2014faces,booth2017toward,kaur2018prediction,chen2019faceengage}. These functional-based methods suffer from the loss of information caused by squashing long sequences of features to functionals (i.e., numbers). A further difficulty is manually determining the appropriate type of functionals (differentiating between different levels of engagement) and the optimal size of windows within which to calculate the functionals.

Aside from the image- and functional-based methods, most end-to-end and feature-based engagement measurement methods utilize spatio-temporal or temporal/sequential models to analyze sequences of the behavioral and affect states and the order in which those states occur. A challenge is raised in the following section regarding the necessity of incorporating the order of the states and using temporal/sequential models to analyze the order of the states for engagement measurement.

\section{Does Order Matter in Engagement Detection?}
\label{sec:is_order_important}
The definition of engagement and how it is annotated in the existing video-based student engagement measurement datasets were investigated by Khan et al. \cite{khan2022inconsistencies}. One significant dimension of engagement annotation is temporal resolution (timescale), the timesteps in which the annotation takes place \cite{khan2022inconsistencies,d2015influence}. Correspondingly, the duration of the video samples in the datasets is equal to the temporal resolution of engagement annotation. There is a wide range of temporal resolutions available in the existing datasets, ranging from a second \cite{zaletelj2017predicting} to 30 minutes \cite{ma2021hierarchical}. In the existing datasets, the video samples did not necessarily represent the entire learning session and were not necessarily extracted from a specific period of time, e.g., the beginning or end of the learning session \cite{khan2022inconsistencies}.

Research on affect dynamics explores the way affect develops and manifests over time \cite{karumbaiah2021re}. According to the extensive experiments on the dynamics of affective states during learning \cite{d2012dynamics,d2007monitoring,d2007dynamics}, there is an affect state transition approximately every 10-40 seconds. Therefore, the video samples in datasets with a low temporal resolution of engagement annotation, e.g., five minutes, may contain more than one affect state or more than one indicator of engagement. Considering a video sample as a multiset or bag of words, multiple indicators of engagement (multiple words) may occur in the timescale in which the video sample has been annotated.

Various models have been developed for affect dynamics describing how students transition from one affect state to the next during learning activities \cite{d2012dynamics,karumbaiah2021re}. For instance, Karumbaiah et al. \cite{karumbaiah2021re} demonstrated that it is significantly likely to transition from an engagement state to a confusion state, but it is unlikely to transition from a frustration state to an engagement state. However, no such model exists for engagement levels \cite{khan2022inconsistencies}. None of the existing definitions of engagement or protocols for engagement annotation have specified that a particular pattern of engagement levels with a specific order must occur in a video sample to annotate it as a specific engagement level. In terms of the indicators of engagement, there is no unified definition such that particular indicators of affective and behavioral engagement or disengagement with a specific order must occur in a video sample to annotate it as a specific engagement level \cite{khan2022inconsistencies,abedi2022detecting}, refer to Figure 1 for an example.

By observing the existing publicly available video-based student engagement datasets \cite{gupta2016daisee,kaur2018prediction,khan2022inconsistencies}, we deduced that engagement and disengagement take on diverse \textit{temporal} manifestations. The behavioral and affective states manifesting engagement and disengagement in a video sample are well understood and well defined; a student who is attentively looking at the camera and exhibiting a focused affect state is considered engaged; a student who is working on their phone and does not focus on the camera and is in a sleepy affect state is considered disengaged. However, the temporal changes in the states and their order is not well defined to consider the student in a video sample as engaged or disengaged. As a result, it is difficult for a temporal/sequential neural network to be trained and successfully model the temporal transitions and order of the states and output the level of engagement. The inability to model the temporal changes in the states presents more challenges for disengaged video samples \cite{abedi2022detecting}. Because in many existing student engagement datasets, the distribution of engaged to disengaged samples is highly imbalanced; in many cases, the percentage of disengaged samples is extremely low \cite{dresvyanskiy2021deep,khan2022inconsistencies,karimah2022automatic}.

As a result, our hypothesis is that the order and position of affective and behavioral states, as well as the pattern of temporal changes in engagement levels, may not be relevant in measuring the engagement level of a student in a video sample. We further hypothesize that only the frequency of occurrence of certain states and engagement levels are important for engagement measurement. Therefore, we consider a video sample as a multiset- or bag-of-words, i.e., a bag of affective and behavioral states. For the first time in the field of engagement measurement, we develop BoW-based models to analyze only the occurrence of affective and behavioral states and not the order of the occurrences. The necessity of incorporating the order and position of words/tokens has also been recently questioned in other fields, such as natural language processing \cite{galke2022bag}. Galke and Scherp \cite{galke2022bag} showed that for text classification, simple BoW-based models, which disregard the order and position of words/tokens, outperform complex graph-based and sequence-based neural network models.

\begin{figure}
    \centering
    \includegraphics[scale=.5]{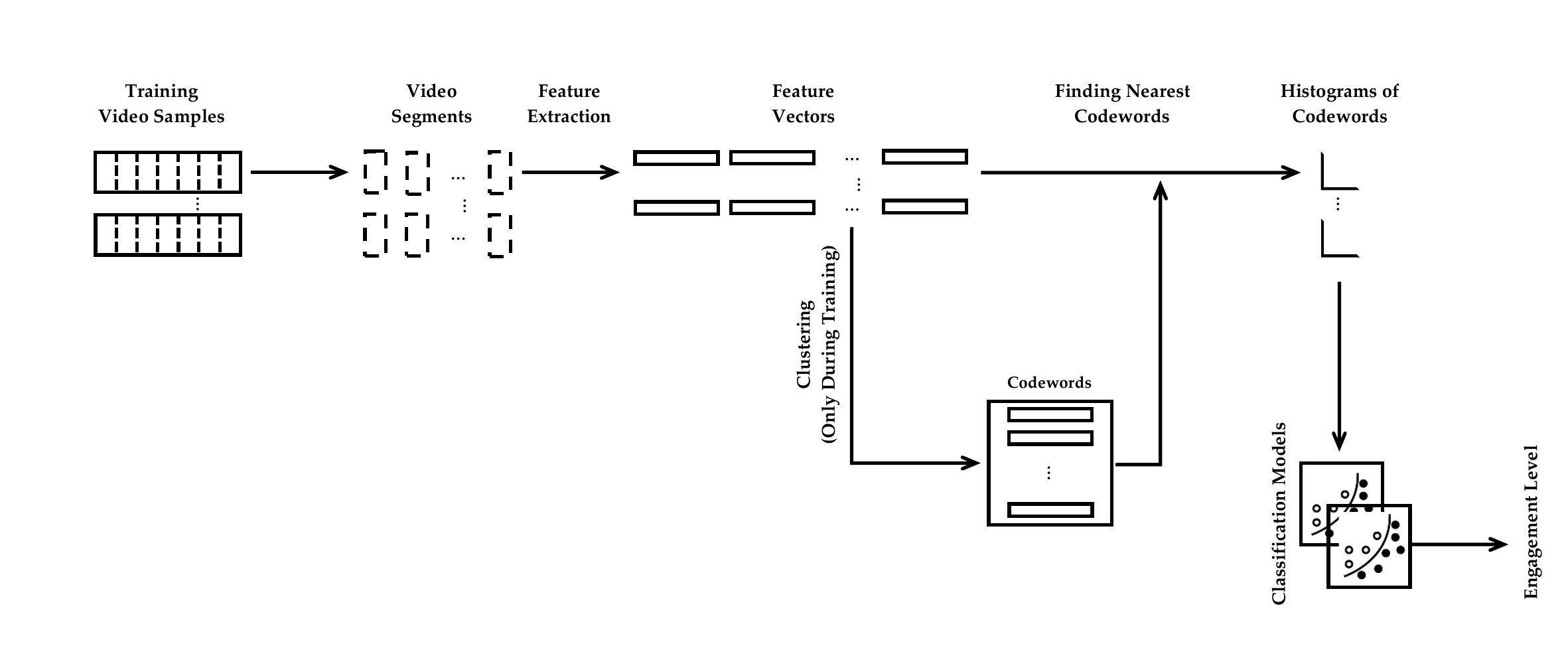}\\
    \caption{The block diagram of the Bag of States (BoS), the proposed non-sequential approach for engagement measurement from videos, based on the bag of words approach, see Section \ref{sec:method}.
    }
    \label{fig:methodology}
\end{figure}

\section{Bag of States}
\label{sec:method}
Figure \ref{fig:methodology} illustrates the block diagram for BoS, the proposed non-sequential approach for engagement measurement from videos. The input is a video sample of a student in a virtual learning session sitting in front of the camera of a laptop or PC (as shown in Figure \ref{fig:frames}), and the output is the engagement level of the student in the video.

In the training phase, all video samples in the training set are divided into video segments of equal length\footnote{In the video engagement measurement datasets we are dealing with, there is one annotation for each video sample indicating the engagement level of the student throughout the whole video sample. The video sample is in the range of seconds to minutes in length. As part of our approach, the video samples are divided into video segments. A noteworthy observation is the absence of engagement annotations for video segments.}.
A variety of behavioral and affect features (see Section \ref{sec:feature_extraction}) are extracted from each video segment representing the behavioral and affective states of the student in the segment\footnote{A video sample is considered to be a multi-set or bag of (behavioral and affective) states.}.

\begin{figure}
    \centering
    \includegraphics[scale=.5]{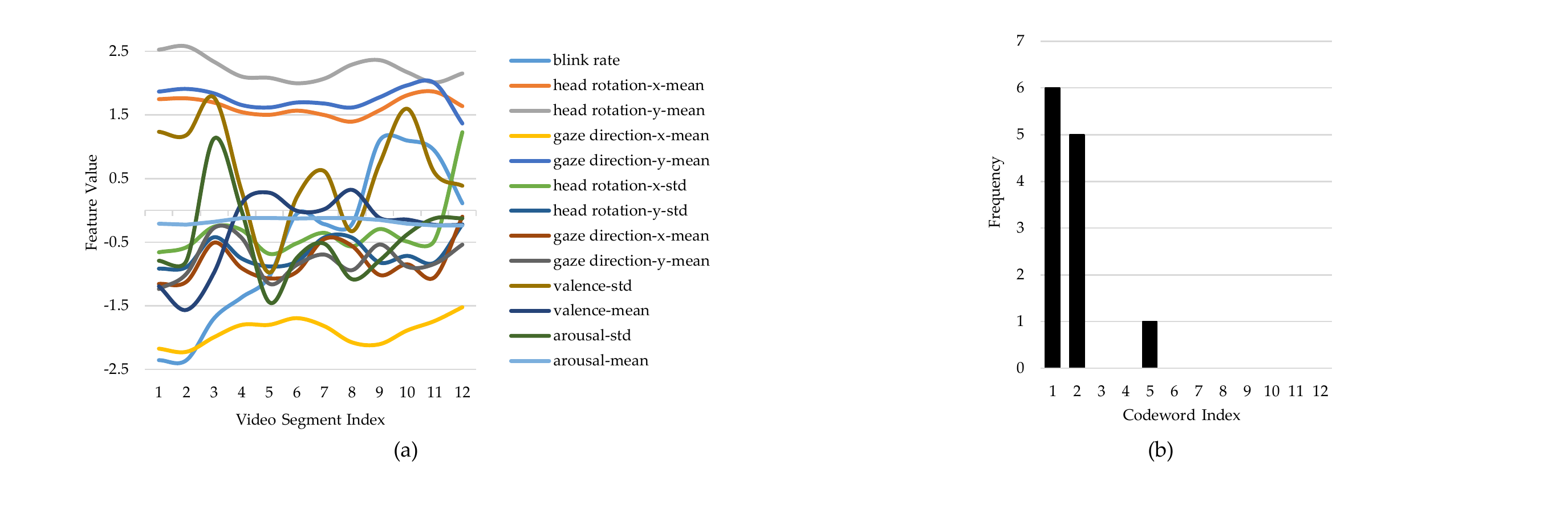}\\

    \caption{
    (a) An exemplary subset of features extracted from 12 video segments of the video sample in Figure \ref{fig:frames} (a), and (b) the generated histogram of 12 codewords for the video sample in Figure \ref{fig:frames} (a), see Section \ref{sec:method}.
    }
    \label{fig:histogram}
\end{figure}

In the next step, following the idea of bag-of-words in text analysis \cite{lebanon2007locally} and bag-of-visual-words in image and video analysis \cite{niebles2008unsupervised}, the feature vectors extracted from video segments are clustered to create a codebook with the cluster centers serving as codewords. A codeword can be viewed as a representative of several similar behavioral and affective states. The number of clusters is the codebook size. The K-means, with Lloyd's algorithm for clustering and K-means++ for cluster center initialization, is used for codebook generation. Each video segment (i.e., its corresponding feature vector) is mapped to a certain codeword through the clustering process, and as a result, each video sample is represented by a histogram of the codewords. Notable, the histograms of the codewords provide information about the frequency of occurrence of codewords in the video samples, not their order of occurrence, refer to Figure \ref{fig:histogram} for an example.

Although temporal versions of BoW models were developed to incorporate the order of occurrence of patterns \cite{bettadapura2013augmenting,govender2020spatio}, due to the fact that our objective is to ignore the temporal order, we adhere to the vanilla histograms of codewords. The generated histograms of codewords (i.e., bag-of-states representations) for video samples and their corresponding engagement level annotations are then used to train supervised machine-learning models for engagement measurement.

In the datasets with multi-level engagement annotations, engagement is an ordinal variable \citep{khan2022inconsistencies,whitehill2014faces}, as opposed to a nominal variable. Therefore, in the BoS approach, the ordinal version of the machine-learning model for the classification of the histograms of codewords is developed as follows. The original $(K + 1)$-level ordinal labels of engagement levels, $y = 0, 1, …, K$ in the training set are converted into $K$ $y_i$ binary labels as follows, if $y > i: y_i = 0$, else: $y_i = 1$, $i = 0, 1, …, K – 1$. Then, $K$ binary classifiers ($C_i$, $i = 0, 1, …, K – 1$) are trained using the histograms of codewords for the training samples and $K$ $y_i$ binary labels.

In the test phase, a test video sample is divided into video segments of equal length, and behavioral and affect features (see Section \ref{sec:feature_extraction}) are extracted from the segments. Each video segment is mapped to a certain codeword which has a minimum euclidean distance to it, and as a result, each video sample is represented by a histogram of the codewords. The generated histogram is fed into the trained ordinal machine-learning model to output the level of engagement as follows. Each trained binary classifier $C_i$ gives a probability estimate for the test histogram $x_t$, the probability of being in binary class $y_i$, $y_t = y_i$, $i = 0, 1, …, K – 1$. Then, $K$ binary probability estimates are converted into one multi-class probability of being in class $y = 0, 1, …, K$ as follows,

\begin{eqnarray}\label{eqexpmuts}
p(y_t=k) =
\left\{
	\begin{array}{ll}
		1-p(y_t>0),  & \mbox{if }  k=0 \\
		p(y_t>k-1)-p(y_t>k), & \mbox{if } 0<k<K-1 \\
		p(y_t>K-2), & \mbox{if } k=K-1
	\end{array}
\right..
\end{eqnarray}

\subsection{Feature Extraction}
\label{sec:feature_extraction}
Various affect and behavioral features are extracted from the video segments \cite{abedi2021affect,abedi2022detecting}. Abedi and Khan \cite{abedi2021affect} showed that sequences of continuous values of valence and arousal extracted from consecutive video frames are important indicators of affective engagement. The pre-trained EmoFAN \cite{toisoul2021estimation} on the AffectNet dataset \citep{mollahosseini2017affectnet}, a deep neural network to analyze facial affects in face images, is used to extract valence and arousal from consecutive frames of video segments (refer to \cite{abedi2021affect} for more detail). Inspired by previous research \cite{abedi2021affect,liao2021deep,thomas2018predicting,kaur2018prediction,niu2018automatic,ranti2020blink}, eye location, head pose, eye gaze direction, blink rate, and hand pose in consecutive frames of video segments are extracted as behavioral features (refer to \cite{abedi2021affect} for more detail). Then, for each video segment, a 49-element segment-level feature vector is created as follows.
\begin{itemize}
  \item 4 features: the mean and standard deviation of valence and arousal values over consecutive clip frames,
  \item 1 feature: the blink rate, derived by counting the number of peaks above a certain threshold divided by the number of frames in the AU45 intensity time series extracted from the input clip,
  \item 8 features:  the mean and standard deviation of the velocity and acceleration of x and y components of eye gaze direction,
  \item 12 features:  the mean and standard deviation of the velocity and acceleration of x, y,  and z components of head location,
  \item 12 features:  the mean and standard deviation of the velocity and acceleration of head’s pitch, yaw, and roll, and
  \item 12 features:  the mean and standard deviation of the velocity and acceleration of x, y,  and z components of wrist location.
 \end{itemize}


\section{Experiments}
\label{sec:experiments}
This section evaluates the BoS engagement measurement method in comparison to the previous feature-based and end-to-end methods. The classification results on two video-based engagement datasets are reported. The necessity of incorporating the order of behavioral and affective states in engagement measurement is investigated through a comparison between the proposed non-sequential BoS approach with sequential approaches. Different feature sets are ablation studied, and the effectiveness of affect states in engagement measurement is evaluated. Furthermore, the impact of incorporating the ordinality of engagement variable in model training is investigated by comparing the results of ordinal-output and non-ordinal-output models.

\subsection{Datasets}
\label{sec:datasets}
The performance of the BoS method is evaluated on our collected video engagement dataset, the IIITB Online Student Engagement (IIITB Online SE) \cite{thomas2022automatic}, as well as a publicly available video engagement dataset, the Dataset for Affective States in E-Environments (DAiSEE) \cite{gupta2016daisee}.

\noindent
\textbf{IIITB Online SE:} The IIITB Online SE dataset \cite{thomas2022automatic} contains videos captured from 6 students attending online courses. The videos were recorded using the Zoom platform. The length, frame rate, and resolution of the videos are 100 seconds, 24 frames per second (fps), and 220 × 155 pixels. The videos of students were self-annotated through self-reports in terms of engagement level. There was, however, only one self-report engagement annotation for each student during the entire learning session. Therefore, using the Human Expert Labeling Process (HELP) definition \cite{aslan2017human}, the 100-second video samples were annotated by external observers in terms of the engagement of the student in the video samples (engaged vs. not-engaged). After removing corrupted video samples, the dataset contains 319, 166, and 129 samples in the train, validation, and test sets, respectively. Table \ref{tab:tab_iiitbonlinese} shows the distribution of samples in engaged and not-engaged classes in the train, validation, and test sets.

\noindent
\textbf{DAiSEE:} The DAiSEE dataset \cite{gupta2016daisee} contains 9,068 videos captured from 112 persons in online courses. The videos were annotated by four states of persons while watching online courses, boredom, confusion, frustration, and engagement. Each state is in one of the four levels (ordinal classes), level 0 (very low), 1 (low), 2 (high), and 3 (very high). In this paper, the focus is only on the engagement level classification. The length, frame rate, and resolution of the videos are 10 seconds, 30 fps, and 640 × 480 pixels. Table \ref{tab:tab_daisee} shows the distribution of samples in different classes and the number of persons in the train, validation, and test sets. As can be seen in Table \ref{tab:tab_daisee}, the dataset is highly imbalanced. One major difference between the above two datasets is the duration of the video samples, 100 seconds in IIITB Online SE and 10 seconds in DAiSEE.


\begin{table}[ht]
\caption{The distribution of samples in engaged and not-engaged classes in the train, validation, and test sets in the IIITB Online SE dataset \cite{thomas2022automatic}.}
\label{tab:tab_iiitbonlinese}
\centering
\begin{tabular}{p{.15\linewidth}p{.15\linewidth}p{.15\linewidth}p{.15\linewidth}}
\hline
level & train & validation & test\\
\hline
not-engaged & 57 & 58 & 57\\
\hline
engaged & 262 & 108 & 72\\
\hline
total & 319 & 166 & 129\\
\hline
\end{tabular}
\end{table}

\begin{table}[ht]
\caption{The distribution of samples in different engagement level classes, and the number of persons, in train, validation, and test sets in the DAiSEE dataset \cite{gupta2016daisee}.}
\label{tab:tab_daisee}
\centering
\begin{tabular}{p{.2\linewidth}p{.15\linewidth}p{.15\linewidth}p{.15\linewidth}}
\hline
level & train & validation & test\\
\hline
0 & 34 & 23 & 4\\
\hline
1 & 213 & 143 & 84\\
\hline
2 & 2617 & 813 & 882\\
\hline
3 & 2494 & 450 & 814\\
\hline
total & 5358 & 1429 & 1784\\
\hline
\# of persons & 70 & 22 & 20\\
\hline
\end{tabular}
\end{table}


\subsection{Evaluation Metrics}
\label{sec:evaluation_metrics}
In accordance with the engagement measurement problem in the two datasets described above, a variety of evaluation metrics are used. For the binary engagement classification problem in the IIITB Online SE dataset \cite{thomas2022automatic}, accuracy, precision, recall, and F1 scores are reported. For the multi-class engagement level classification problem in the DAiSEE dataset \cite{gupta2016daisee}, accuracy is reported.

\subsection{Experimental Results}
\label{sec:experimental_results}
The behavioral eye and head features, and hand features (described in Section \ref{sec:feature_extraction}) are extracted by the OpenFace \cite{baltrusaitis2018openface}, and MediaPipe \cite{lugaresi2019mediapipe}, respectively. The OpenFace also outputs the extracted face regions from video frames. The extracted face regions are fed to the pre-trained EmoFAN \cite{toisoul2021estimation} on AffectNet \cite{mollahosseini2017affectnet} for extracting affect features \ref{sec:feature_extraction}. The machine-learning and deep-learning modeling were implemented in PyTorch \cite{paszke2019pytorch} and Scikit-learn \cite{pedregosa2011scikit} on a server with 64 GB of RAM and NVIDIA TeslaP100 PCIe 12 GB GPU.

\subsubsection{Comparison with Sequential Methods}
\label{sec:sequential_methods}
The importance of the features outlined in Section \ref{sec:feature_extraction} was explored by Abedi and Khan in \cite{abedi2021affect}. It has been demonstrated that the combination of behavioral and affect features performs better than either alone. As part of this study, the performance of the BoS was also evaluated using different feature combinations. To investigate the necessity of modeling the order of the states in engagement measurement, BoW modeling in BoS was compared with TCN, the sequential model achieving the state-of-the-art results in the previous works on engagement measurement \cite{thomas2022automatic,thomas2018predicting,abedi2021improving}. Table \ref{tab:tab_comparison_sequential} shows the results of the BoW modeling in BoS in comparison with TCN using different feature sets on the test set of the IIITB Online SE dataset.

According to Table \ref{tab:tab_comparison_sequential}, BoW and TCN models both perform better with the behavioral feature set than with the affect feature set alone and adding affect features to behavioral features also results in significant improvement. This result can be interpreted in accordance with the psychological definition of engagement (Section \ref{sec:introduction}) which encompasses the behavioral and emotional states of the student. There is a significant difference between the BoW and TCN models in all the three feature sets shown in Table \ref{tab:tab_comparison_sequential}, indicating that a non-sequential model ignoring the order of behavioral and affect states is superior to a sequential model that attempts to model the order of states. It is noteworthy that this result is consistent with the discussion in Section \ref{sec:is_order_important}.

In the experiments in Table \ref{tab:tab_comparison_sequential}, the size of the segments the video samples were divided into was 200 frames, giving the best results on the validation set. The 100-second (2400 frames) video samples were divided into 12 segments, 200 frames each. The behavioral and affect feature vectors extracted from consecutive video segments were fed to consecutive timestamps of the TCN. The parameters of the TCN in Table \ref{tab:tab_comparison_sequential}, giving the best results, are as follows, 3, 25, 7, and 0.2, 64 for the number of levels, number of hidden units, kernel size, dropout, and batch size \cite{bai2018empirical}. The TCN is trailed by a fully-connected layer with one output neuron and sigmoid for binary engagement classification.

\begin{table}[ht]
\caption{The results of the BoW modeling in the proposed BoS engagement measurement method in comparison with TCN modeling using different Behavioural (B) and/or Affect (A) feature sets on the test set of the IIITB Online SE dataset.
}
\label{tab:tab_comparison_sequential}
\centering
\begin{tabular}{p{.125\linewidth}p{.1\linewidth}p{.1\linewidth}p{.1\linewidth}p{.1\linewidth}p{.1\linewidth}}
\hline
Features & Model & Accuracy & Precision & Recall & F1 Score\\
\hline
B & TCN & 0.71 & 0.67 & 0.90 & 0.77\\
\hline
B & BoW & 0.86 & 0.81 & 0.96 & 0.88\\
\hline
A & TCN & 0.54 & 0.54 & 1.0 & 0.70\\
\hline
A & BoW & 0.65 & 0.61 & 1.0 & 0.76\\
\hline
B + A & TCN & 0.74 & 0.70 & 0.91 & 0.79\\
\hline
\textbf{B + A} & \textbf{BoW} & \textbf{0.93} & \textbf{0.93} & \textbf{0.91} & \textbf{0.95}\\
\hline
\end{tabular}
\end{table}

\subsubsection{Parameters of the Bag-of-States Method}
\label{sec:model_paramters}
\begin{figure}
    \centering
    \includegraphics[page=1,scale=.5]{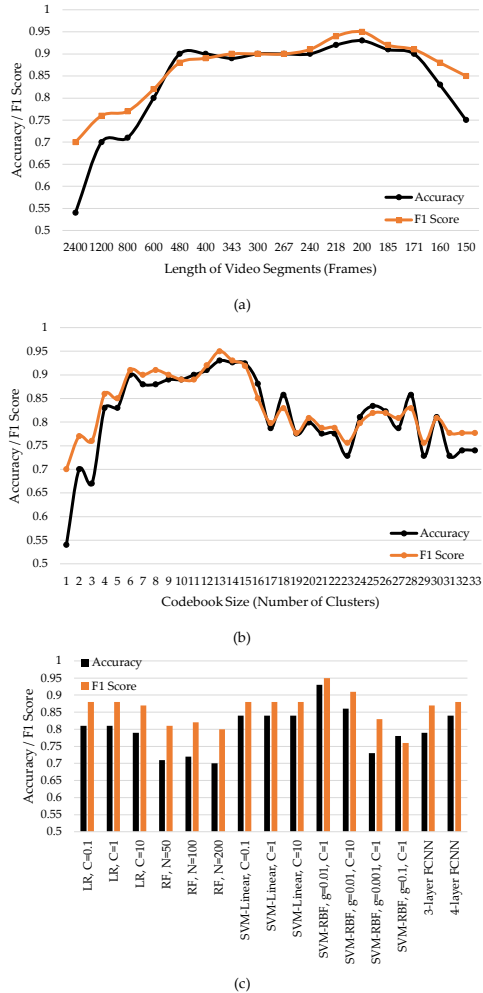}\\

    \caption{
    The performance of the BoS method on the IIITB Online SE dataset using different (a) sizes of video segments (in frames), (b) codebook sizes, and (c) models for classification of histograms of codewords, LR: Logistic Regression, RF: Random Forest, SVM: Support Vector Machine,  RBF: Radial Basis Function, FCNN: Fully-Connected Neural Network.
    }
    \label{fig:parameters}
\end{figure}

The BoS method has two main parameters, the size of the segments the video samples are divided into and the codebook size, i.e., the number of clusters into which the training video segments are clustered. In addition, various classification models can be used for the classification of the histograms of codewords and outputting the engagement level. For the experiments in Table \ref{tab:tab_comparison_sequential}, giving the best results on the validation set, the size of the segments, and the codebook size were 200 frames, and 12, respectively. SVM with Radial Basis Function (RBF) kernel, C = 1.0, and gamma = 0.01 were used for classification. Considering that the engagement measurement in the IIITB Online SE dataset is a binary classification problem, there is no need to implement the ordinal version of the classifier. Figure \ref{fig:parameters} shows the effects of changing single parameters of the BoS method while other parameters remain unchanged, as described in the above-described setting.

In Figure \ref{fig:parameters} (a), the size of video segments (in frames) was changed, while the values of other parameters were left unchanged, and accuracy and F1 score were reported. Starting from video segments of size 2400 frames, i.e., considering the entire video samples as segments, the accuracy was around 0.5, a random binary classifier. It is difficult to represent variations in behavioral and affect states using 12 codewords in such long video segments. Moreover, the calculation of mean and standard deviation functionals for feature extraction in long video segments results in the loss of information. The performance was steady with segments with a length of 160 to 480 frames, and the best results were achieved with segments of 200 frames. From an engagement measurement perspective, it was observed that the micro-level indicators of engagement/disengagement can be best measured in the time resolutions of length 160 to 480 frames (6 to 20 seconds).

In Figure \ref{fig:parameters} (b), the codebook size was changed, while the values of other parameters were left unchanged, and accuracy and F1 score were reported. Under-fitting was caused by a small number of codewords (less than four), and over-fitting was caused by a large number of codewords (more than 11). Feature vectors extracted from video segments cannot be represented by a small number of codewords, for example, by a histogram of codewords with fewer than four bins. A histogram of codewords with a large number of bins tends to model small changes and noise in the extracted feature vectors from video segments. A further consideration is that the optimal number of codewords depends on the size of the video segments. Longer video segments require larger numbers of codewords to be represented, and vice versa.

Another experiment in Figure \ref{fig:parameters} (c) examined the use of different models to classify histograms of codewords and determine engagement levels. LR with different regularization parameters, RF with different trees, SVM with linear kernel and different regularization parameters, SVM with RBF kernel and different regularization parameters and kernel coefficients, and fully-connected neural networks with Adam optimizer and different numbers of layers. As can be seen in Figure \ref{fig:parameters} (c), the best results were obtained by the SVM with RBF kernel having kernel function of 0.01 and regularization parameter of 1.

\subsubsection{Comparison with Sequential Methods using Self-report annotations}
\label{sec:self_reports}
As explained in Section \ref{sec:datasets}, the IIITB Online SE dataset was re-annotated by external observers using the HELP definition of engagement \cite{aslan2017human} and used for the experiments throughout this paper. As a means of comparison between the proposed method and the previous work by the creators of the dataset \cite{thomas2022automatic}, in one experiment in Table \ref{tab:self_reports}, self-reports were used as the annotations of the samples. The results of the proposed non-sequential method are marginally superior to the two methods presented in \cite{thomas2022automatic}. In the IIITB Online SE dataset, there is only one student self-report annotation for the entire virtual learning session. The authors propagated the session-level self-report annotation for the 100-second video samples extracted from the video of the session. This is problematic since the level of student engagement is a state that may not maintain a steady level for more than a minute \cite{d2012dynamics}. Thus, there are different levels of engagement throughout the entire session, and a single annotation can be misleading to machine-learning models attempting to model engagement.

\begin{table}[ht]
\caption{The results of the BoS engagement measurement method on the IIITB Online SE dataset with self-report annotations compared to the previous work \cite{thomas2022automatic}.
}
\label{tab:self_reports}
\centering
\begin{tabular}{p{.3\linewidth}p{.075\linewidth}p{.075\linewidth}p{.075\linewidth}p{.075\linewidth}p{.1\linewidth}}
\hline
Features & Model & Accuracy & Precision & Recall & F1 Score\\
\hline
Facial Action Units \cite{thomas2022automatic,baltrusaitis2018openface} & TCN & 0.74 & 0.76 & 0.85 & 0.80\\
\hline
MIMAMO \cite{thomas2022automatic,Deng_Chen_Zhou_Shi_2020} & TCN & 0.76 & 0.77 & 0.89 & 0.82\\
\hline
\textbf{Behavioral and Affect} & \textbf{BoW} & \textbf{0.79} & \textbf{0.74} & \textbf{0.93} & \textbf{0.82}\\
\hline
\end{tabular}
\end{table}

\subsubsection{Comparison with Previous Works}
\label{sec:comparison_previous_works}
Table \ref{tab:daisee_results_1} shows the results of the proposed method compared to the previous methods on the test set of the DAiSEE dataset. Since engagement level is a 4-level ordinal variable in the DAiSEE dataset, the ordinal version of the BoS model is implemented. These are the parameters of the proposed method for the DAiSEE dataset giving the best results on the validation set: The video segments are 75 frames long, and the codebook size is 8. Ordinal classification is performed using SVM with an RBF kernel with a regularization parameter of 1 and a kernel function of 0.01.

Most of the end-to-end and feature-based methods outlined in Table \ref{tab:daisee_results_1} are complex deep-learning sequential models incorporating the order of the behavioral and affect states in engagement measurement. The proposed lightweight BoS model, ignoring the order of events, achieved an accuracy of 66.58\%, one of the highest accuracies among the existing works. Only the accuracies of two complex methods are above the proposed method, Hybrid EfficientNetB7 with LSTM \cite{selim2022students}, and affect-driven ordinal TCN \cite{abedi2021affect}. By comparing the proposed method to previous methods, it becomes clear that incorporating the order of states in engagement measurement may not be unnecessary. In addition, the results of methods based on functionals are presented in Table \ref{tab:daisee_results_1}. In these methods, the features introduced in Section \ref{sec:feature_extraction} are extracted from the entire video samples and then classified by a traditional machine-learning model. The results of methods based on functionals are significantly lower than those of sequential and BoS models.

\begin{table}[ht]
\caption{Engagement level classification accuracy of the proposed method compared to the previous methods on the DAiSEE dataset.}
\label{tab:daisee_results_1}
\centering
\begin{tabular}{p{.85\linewidth}p{.1\linewidth}}
\hline
method & accuracy\\
\hline
C3D \cite{gupta2016daisee} & 48.10\\
\hline
I3D \cite{zhang2019novel} & 52.40\\
\hline
C3D + LSTM \cite{abedi2021improving} & 56.60\\
\hline
C3D with transfer learning \cite{gupta2016daisee} & 57.80\\
\hline
LRCN \cite{gupta2016daisee} & 57.90\\
\hline
DFSTN \cite{liao2021deep} & 58.80\\
\hline
C3D + TCN \cite{abedi2021improving} & 59.90\\
\hline
DERN \cite{huang2019fine} & 60.00\\
\hline
ResNet + LSTM \cite{abedi2021improving} & 61.50\\
\hline
Neural Turing Machine \cite{ma2021automatic} & 61.30\\
\hline
3D DenseNet \cite{mehta2022three} & 63.60\\
\hline
ResNet + TCN \cite{abedi2021improving} & 63.90\\
\hline
ShuffleNet v2 \cite{hu2022optimized} & 63.90\\
\hline
EfficientNetB7 + TCN \cite{selim2022students} & 64.67\\
\hline
Affect-driven Ordinal TCN  \cite{abedi2021affect} & 67.40\\
\hline
EfficientNetB7 + LSTM \cite{selim2022students} & 67.48\\
\hline
\hline
Functional features + SVM & 55.75\\
\hline
Functional features + RF & 59.09\\
\hline
\hline
BoS, behavioral and affect features, non-ordinal output classifier (proposed) & 65.43\\
\hline
BoS, behavioral, ordinal-output classifier (proposed) & 64.84\\
\hline
BoS, behavioral and affect, ordinal-output classifier (proposed) & 66.58\\
\hline
\end{tabular}
\end{table}

\section{Conclusion}
\label{sec:conclusion}
In this paper, we proposed a novel method for engagement measurement from videos of students in online courses. According to the definition of engagement in educational psychology \cite{sinatra2015challenges,woolf2009affect}, it can be measured at fine-grained time scales through the analysis of the behavioral and affective states of students. The combination of certain behaviors and affect states, such as on-task/off-task behavior and positive/negative valence and arousal values, may serve as indicators of engagement or disengagement. The definitions of engagement do not specify that behavioral and affect states must occur in a particular order to consider the student engaged or not-engaged. Nevertheless, almost all the existing works on video-based engagement measurement developed sequential models and tried to model the sequences and temporal changes in the behavioral and affect states to measure engagement. In this paper, backed by educational psychology, it was hypothesized that incorporating the order of the states may not be necessary for engagement measurement; nonetheless, their occurrences are important to be considered as indicators of engagement/disengagement. A video sample for engagement measurement was considered as a multiset (bag) of behavioral and affect states. A non-sequential model based on BoW, BoS, was developed for engagement measurement. The results of experiments conducted on two student engagement measurement datasets \cite{thomas2022automatic,gupta2016daisee} confirmed the strength of our hypothesis. Most of the time, the lightweight BoS method performed better than complex deep-learning sequential models. The BoS method is novel in the sense that it is the first work in the field of engagement measurement based on BoW operating without incorporating the temporal order of states. In order to develop a simple yet effective method without incorporating the order of states, we adopted the traditional version of the BoW with modifications.

One limitation of the BoS method is its inability to handle variable-length videos. In the current setting, the BoS method extracts larger numbers of video segments from longer videos. Therefore there will be larger numbers of occurrences of certain behavioral and affect states which are indicators of engagement/disengagement and could be misleading for the BoS model. In future work, we will work on enabling BoS to handle variable-length sequences. Another extension of BoS we plan to explore is modifying BoS for anomaly detection in affective computing problems, disengagement detection in our case \cite{abedi2022detecting}. We will investigate how BoS can be applied across datasets, i.e., creating a codebook from one dataset and using it to generate histograms of codewords and engagement measurements in another dataset. In another line of investigation, we will not only compare the performance of engagement measurement using the BoS approach versus the sequential approach on large datasets, but we will also examine the computational complexity and the time required for inference of each approach.


\bibliography{sn-bibliography}

\end{document}